\documentclass[peerreview]{IEEEtran}
\IEEEoverridecommandlockouts
\usepackage{cite}
\usepackage{amsmath,amssymb,amsfonts}
\usepackage{algorithmic}
\usepackage[ruled,vlined]{algorithm2e}
\usepackage{graphicx, caption, subcaption}
\usepackage{multirow, multicol, makecell, booktabs}
\usepackage{textcomp}
\usepackage[table,xcdraw]{xcolor}
\usepackage{xcolor}
\usepackage{pdflscape}
\usepackage{afterpage}
\usepackage{enumitem}
\newlist{steps}{enumerate}{1}
\setlist[steps, 1]{label = Step \arabic*:}

\def\BibTeX{{\rm B\kern-.05em{\sc i\kern-.025em b}\kern-.08em
    T\kern-.1667em\lower.7ex\hbox{E}\kern-.125emX}}
\begin{document}

\title {Robust Black-box Watermarking for Deep Neural Network using Inverse Document Frequency \\
}


\author{\IEEEauthorblockN{Mohammad Mehdi Yadollahi, Farzaneh Shoeleh, Sajjad Dadkhah, Ali A. Ghorbani}
\IEEEauthorblockA{\textit{Canadian Institute for Cybersecurity} \\
\textit{University of New Brunswick}\\
Fredericton, Canada \\
\{m.yadollahi, fshoeleh, sdadkhah, ghorbani\}@unb.ca}
}

\maketitle

\begin{abstract}

Deep learning techniques are one of the most significant elements of any Artificial Intelligence (AI) services. Recently, these Machine Learning (ML) methods, such as Deep Neural Networks (DNNs), presented exceptional achievement in implementing human-level capabilities for various predicaments, such as Natural Processing Language (NLP), voice recognition, and image processing, etc. Training these models are expensive in terms of computational power and the existence of enough labeled data. Thus, ML-based models such as DNNs establish genuine business value and intellectual property (IP) for their owners. Therefore the trained models need to be protected from any adversary attacks such as illegal redistribution, reproducing, and derivation. Watermarking can be considered as an effective technique for securing a DNN model. However, so far, most of the watermarking algorithm focuses on watermarking the DNN by adding noise to an image. To this end, we propose a framework for watermarking a DNN model designed for textual domain. The watermark generation scheme provides a secure watermarking method by combining  Term Frequency (TF) and Inverse Document Frequency (IDF) of a particular word. The proposed embedding procedure takes place in the model's training time, which makes the watermark verification stage straightforward by sending the watermarked document to the trained model. The experimental results show that watermarked models have the same accuracy as the original one, and the proposed framework accurately verifies the ownership of all surrogate models without impairing the performance. The proposed algorithm is robust against well-known attacks such as parameter pruning and brute force attack.

\end{abstract}



\section{Introduction}


Deep learning is a variety of machine learning structure that automatically extracts features from training data to perform better on many complicated tasks. Deep learning techniques produce better results when combines with other machine learning approaches such as neural networks. The combination of deep learning and neural networks are called deep neural networks (DNNs). Lately, DNN algorithms have obtained state-of-the-art results in multiple ML related domains such as natural processing language, voice recognition, and computer vision.



The most important factor that helps DNN to achieve such outstanding results is leveraging numerous labeled training data.
Based on the 2016 data science report\cite{crowd2016}, most data scientists spend 80\% of their time collecting, preparing, and managing data. Since collecting a large amount of labeled data and providing powerful hardware for training a DNN can be very expensive, most scientists prefer to employ a pre-trained model for their problems. On the other hand, organizations consider a trained DNN as their Intellectual Property (IP) because of the costs of collecting a considerable amount of labeled data and providing powerful hardware for training it.

Pre-trained models help users to develop their specific methods, which is called fine-tuning. Fine-tuning obtains the best set of weights for a trained DNN in a particular problem and uses them as initialization weights for a new model in the same domain. Fine-tuning is an effective technique to speed up the training phase of DNN models, and also it helps overcome the small dataset problem. Fine-tuning assists scientists to build an accurate and high-performance DNN model. A typical adversary can utilize different fine-tuning approaches as a means for redistribution and copyright infringement. Thus, a trained DNN model needs to be protected as an Intellectual Property (IP) from illegal redistribution, reproducing, and derivation. Digital Watermarking is one of the best solutions to protect a trained DNN model from copyright infringement.


Digital watermarking is a technique that embeds different types of data, signal, or information into digital media such as digital image, audio, and video \cite{van1994digital}. Digital watermarking can have several different purposes. These include merely hiding a piece of meaningful information without modifying the host of the watermark or embedding a piece of specific information that can ensure the originality of a digital file by authenticating the embedding content.


Various watermarking algorithms utilize several different approaches to distinguish between original or watermarked media. The watermarking algorithms encrypt the content using various encryption techniques such as block ciphers to avoid revealing the watermark's information to adversaries who have prior knowledge of the watermarking algorithm. Recently, many approaches have been published to watermark the DNNs to protect them from abuse cases. In these methods, the owner of a trained DNN model watermarks his/her model by embedding specific data into the training dataset or modifying some parameters of the model.


To the best of our knowledge, all methods and strategies proposed for watermarking DNNs are focused and evaluated on digital image classification tasks because adding noise to a digital image within a dataset can be very straightforward. In this research, we propose a framework to watermark a DNN model that is trained with textual data. The first stage of the proposed algorithm includes selecting a random data from the training set and adding a certain amount of random noise to the selected data. This set of data is called the \textit{trigger set} and considered as the watermark. After generating the trigger set, the model is trained with a specific combination of this set and original training data. At this step, the trained model is watermarked, which means it returns correct prediction to the ordinary data while it returns a modified response to the trigger set data.



The rest of this paper is organized as follows. Section \ref{RelatedWork} summarizes the important related work that have been proposed to protect DNN models and discusses the related research on watermarking DNN models. In Section \ref{ProposedMethod}, the proposed method for watermarking a textual DNN model is described in detail. The experimental results are presented in Section \ref{ExperimentResults}. Section \ref{Conclusion} provides some concluding remarks and discusses future works.

\section{Related Work} \label{RelatedWork}

In the literature, researchers embed a watermark into a DNN model in three phases: training, fine-tuning, and distillation\cite{adi2018turning}. So, all methods that utilize watermarking to protect a specific DNN are divided into the following three categories:


\begin{itemize}
    \item Watermarking the training data.
    \item Watermarking neural network's parameters.
    \item Watermarking trained model's output.
\end{itemize}

The following sections describe the workflow of the above categories by reviewing the state-of-the-art methods in each section.


\subsection{Watermarking the training data}

The first category of watermarking DNNs refers to those who embed a signature into training data. Zhang et al. \cite{zhang2018protecting} proposed three algorithms for generating digital watermarks as a fingerprint for ownership verification; 1) embedding meaningful content in, for example, inserting a string “TEST” into a car image and labeling it as an airplane. 2) selecting a set of samples from a trivial task, for example, using a handwritten image in the food recognition task. 3) adding a crafted, random, and meaningless noise to some samples. Their method embeds the mentioned watermarks into a target DNNs during the training process and makes the model memorize the watermarks' patterns. Thus, once a DNN is stolen, the owner can easily verify them by sending watermarks as inputs and checking the service output. Their watermarking method was evaluated on two well-known image datasets, MNIST \cite{lecun1998gradient} and CIFAR10 \cite{krizhevsky2009learning}. 
Adi et al. \cite{adi2018turning} proposed a simple yet effective method for watermarking DNNs by employing a backdoor. They declared that for watermarking a machine learning model, the following three steps are essential: generating the secret marking key, embedding the watermark into a model, and verifying if a watermark is present in a model or not. However, they utilized multiple trigger set and generated a random bit string for each of them to insert them into the samples and create a backdoor module. In the verification step, they checked whether the predicted labels are the same as watermarked labels. The quality of their framework is evaluated by estimating non-trivial ownership, residence to removability, and functionality preserving.


Guo et al. \cite{guo2018watermarking} proposed a method to watermark DNNs by adding the author's signature into the portion of the training dataset and assigning them different labels. 
In the watermark detection and verification stage, they run models on samples, both with and without the signature. If the watermarked model classifies original images correctly and classifies images with n-bit string to the mapped classes, they prove that this model is unquestionably their own watermarked model. They used different evaluation criteria such as effectiveness, fidelity, payload, and false-positive rate. They also claimed that their technique could be applied to multiple datasets and DNN architectures, and their model is robust against ghost signature attack and tampering attack.


Huili Chen et al. \cite{chen2019blackmarks} introduced an approach to watermark a pre-trained model in the black-box scenario. Their approach has two main steps; watermark embedding and watermark extraction. In the watermark embedding step, they only have API access. The author's signature is a binary string that all bits are independent of each other. This process includes two main actions, namely, watermark keys generation and fine-tuning. In the watermark extraction step, the owner queries the DNN with watermark keys and decodes the signature from output results. The decrypted mark and actual secret key are compared together for determining the authorship.


Rouhani et al. \cite{rouhani2018deepsigns} proposed a framework to insert a digital watermark in a trained deep learning model. They also introduced reliability and integrity as new requirements of a watermarking framework. For watermarking a DNN, they introduced two approaches: 1) selecting specific target classes and a subset of their training data, and 2) watermarking the output layer by generating a set of unique random input samples and fine-tuning the target model. Their experimental results showed that their approach satisfied all the watermarking requirements and can resist model pruning, fine-tuning, and watermark-overwriting.

\afterpage{%
    \clearpage
    \thispagestyle{empty}
\begin{landscape}
\begin{table}[t]
\centering
\caption{Some of the current deep neural network watermarking methods.}
\label{lit_table}
\scriptsize
\begin{tabular}{|p{2.2cm}|p{5cm}p{5.2cm}p{5cm}p{1.1cm}p{1.1cm}p{1cm}|}
\hline
 &      &     &         &   Access    &   Access    &   Access   \\
Category &   Algorithm   &   Robustness against &   Evaluation metric     &    Model   &    Model   &    Training  \\
 &      &  &   &   Architecture  &  Parameters  &  Data \\\hline
Watermarking the training data
&   Turning your weakness into a strength: Watermarking deep neural networks by backdooring \cite{adi2018turning}   &   Model fine-tuning  &   Model accuracy (0-1)  &   Not \newline Applicable   &   Black-box &   Applicable \\ \cline{2-7} 
    &   Watermarking deep neural networks for embedded systems  \cite{guo2018watermarking}   &   Model fine-tuning  &   Effectiveness, Fidelity and payload with regard to embedding watermarks, and false positive rate with regard to decoding watermarks.  &   Not \newline Applicable   &   Black-box &   Applicable \\ \cline{2-7} 
    &   BlackMarks: Blackbox Multibit Watermarking for Deep Neural Networks \cite{chen2019blackmarks}    &   Brute-force, parameter pruning    &   Accuracy, BER, Detection Success, Overwriting attacks     &   Applicable  &   White-box &   Applicable \\ \cline{2-7} 
    &   Deepsigns: A generic watermarking framework for ip protection of deep learning models \cite{rouhani2018deepsigns}  &   Model fine-tuning, Parameter pruning, Watermark overwriting, lossy compression, cropping, resizing,     &   Accuracy of Marked and Baseline Model     &   Applicable  &   Black-box, White-box &   Applicable \\ \cline{2-7} 
    &   Embedding watermarks into deep neural networks  \cite{uchida2017embedding}   &   Model compression, fine-tuning and distilling    &   Test error (\%) and embedding loss ER(w) with and without embedding, Test error (\%) and embedding loss ER(w) with andwithout embedding in fine-tuning and distilling  &   Not \newline Applicable   &   Black-box &   Applicable \\ \cline{2-7} 
    &   Black-Box Watermarking for Generative Adversarial Networks\cite{skripniuk2020black}    &   Deepfakes and responsibility tracking of GAN misuse, Backdoor attacks, perturbation attacks  &   Thresholding on the bitwise accuracy, FID comparisons    &   Not \newline Applicable   &   Black-box &   Applicable \\ \cline{2-7} 
    &   Watermarking Deep Neural Networks in Image Processing\cite{quan2020watermarking}  &   Compression attacks, Model fine-tuning &   PSNR (dB)/WPSNR  &   Not \newline Applicable   &   Black-box &   Applicable \\ \cline{2-7} 
    &   Evolutionary Trigger Set Generation for DNN Black-Box Watermarking  \cite{guo2019evolutionary}   &   Fine-tune attacks  &   The Key and Logo trigger pattern on different datasets   &   Not \newline Applicable   &   Black-box &   Applicable \\ \cline{2-7} 
    &   Entangled Watermarks as a Defense against Model Extraction\cite{jia2020entangled} &   Retraining-based extraction attacks    &   validation accuracy and watermark success rates(based on cross-entropy of watermarks with target class)    &   Applicable  &   White-box &   Applicable \\ \cline{2-7} 
    &   Training DNN Model with Secret Key for Model Protection \cite{aprilpyone2020training}     &   Brute-force and Fine-tune attacks &   Image classification experiments with a batch size of 128 and live augmentation   &   Not \newline Applicable   &   Black-box &   Not \newline Applicable  \\ \cline{2-7} 
    &   Piracy Resistant Watermarks for Deep Neural Networks \cite{li2019piracy}     &   Piracy Resistant,Corruption, Takeover  &   Normal classification accuracy and watermark accuracies when adversary tries to embed a pirate watermark into owner’s model    &   Not \newline Applicable   &   Black-box &   Not \newline Applicable  \\ \hline
    
Watermarking the training data and 
&   Protecting intellectual property of deep neural networks with watermarking  \cite{zhang2018protecting}   &   Brute-force attacks,model inversion attack, counter-watermark attacks     &   Testing and watermarking accuracy based on different pruning rates &   Applicable  &   Black-box, White-box &   Applicable \\ \cline{2-7} 
 NN’s parameters   &   Digital watermarking for deep neural networks \cite{nagai2018digital} &   Distillation attack, nst parameter pru, Model fine-tuning, lossy compression, cropping, resizing,  &   Test error, Embedding loss, Bit error rate     &   Applicable  &   Black-box &   Applicable \\ \cline{2-7} 
    &   Rethinking Deep Neural Network Ownership Verification: Embedding Passports to Defeat Ambiguity Attacks \cite{fan2019rethinking}     &   network modifications and resilient to ambiguity attacks,fine-tuned, Model pruning  &   Detection/Classification accuracy (in \%) of different passport networks where BN = batch normalization and GN = group normalization     &   Applicable  &   Black-box, White-box &   Applicable \\ \cline{2-7} 
    &   DeepStego: Protecting the Intellectual Property of Deep Neural Networks by Steganography  \cite{zhang2018protecting}   &   Brute-force attacks, model inversion attack, counter-watermark attacks     &   Testing and watermarking accuracy based on different pruning rates &   Applicable  &   Black-box, White-box &   Applicable \\ \hline
    
Watermarking NN’s parameters
&   DeepMarks: A Secure Fingerprinting Framework for Digital Rights Management of Deep Learning Models\cite{chen2019deepmarks}     &   model fine-tuning, parameter pruning, fingerprint collusion, and fingerprint overwriting attacks   &   BIBD AND-ACC codebook that accommodates users, Fine-tune without and with fingerprint  &   Applicable  &   White-box &   Applicable \\ \cline{2-7} 
    &   Robust Watermarking of Neural Network with Exponential Weighting  \cite{namba2019robust}  &   Query modification,     &   Test accuracy of models without watermarks and models watermarked by existing and proposed methods under watermark invalidation in four datasets.  &   Not \newline Applicable   &   Black-box &   Applicable \\ \cline{2-7} 
    &   Adversarial frontier stitching for remote neural network watermarking  \cite{le2020adversarial}     &   Model compression (via both pruning and singular value decomposition) and overwriting via fine-tuning   &   Accuracy with respect to different pruning rates    &   Applicable  &   White-box &   Applicable \\ \cline{2-7} 
    &   Robust and Undetectable White-Box Watermarks for Deep Neural Networks  \cite{wang2019robust}   &   Model fine-tuning , parameter pruning, watermarkoverwriting, property inference attack   &   Accuracy Confidence Intervals and Embedding Loss    &   Applicable  &   White-box &   Applicable \\ \hline
    
Watermarking trained model’s output and NN’s parameters
&   Deep Neural Network Fingerprinting by Conferrable Adversarial Examples \cite{lukas2019deep}    &   Distillation attacks,fine-tuning, ensemble attacks, adversarial trainingand stronger adaptive attacks   &   fingerprint accuracy and fingerprint retention (verification accuracy)  &   Not \newline Applicable   &   Black-box &   Applicable \\ \hline
Watermarking trained model’s output
&   DAWN: Dynamic Adversarial Watermarking of Neural Networks \cite{szyller2019dawn}  &   Model extraction Attacks (IP Theft, KnockOff), Poisoning     &   Accuracy with respect for different epocs &   Not \newline Applicable   &   Black-box &   Not \newline Applicable  \\ \cline{2-7} 
    &   Watermarking the outputs of structured prediction with an application in statistical machine translation  \cite{venugopal2011watermarking}  &   local editing operations     &   Baseline method , Rank interpolation, Cost interpolation, BLEU loss     &   Applicable  &   White-box &   Not \newline Applicable  \\ \hline

\end{tabular}
\end{table}
\end{landscape}
    \clearpage
}


\subsection{Watermark neural network's parameters}

This category of watermarking approach is focused on the structure of DNNs by modifying the parameters of a specific layer in a neural network. These approaches need to have white-access to the neural network. Nagai et al. \cite{nagai2018digital} introduced a digital watermarking technology for authorship authentication of DNNs. They embedded a watermark into the model in three different situations: training, fine-tuning, and distilling situation. They formulated watermarking as embedding a T-bit vector as a secret key in one or more layers of a neural network. The secret key is generated in three different ways; direct, difference and random. The main difference between these three ways is how they choose parameters and layers for modification. By using the secret key, they could embed and detect a watermark in a DNN. 


Based on Huili Chen et al. \cite{chen2019deepmarks}, the two main requirements for copyright protection techniques include ownership proof and tracking individual users. Watermarking techniques and fingerprints are usually a proper solution for copyright protection cases. However, watermarking satisfied the first requirement; in contrast, the fingerprinting can simultaneously address both prerequisites. They proposed a fingerprinting framework named \textit{DeepMarks} that generates a binary code vector for each user and then embedding it in the one or more layer's parameters. Their approach has two steps; 1) generating a unique fingerprint for each user, and 2) inserting the user's signature in selected layers of a pre-trained model by using a secret matrix. For fingerprint extraction, they extracted the code vector from the model and compare it to the codebook column. They claimed that \textit{DeepMarks} is robust against parameter pruning, model fine-tuning, fingerprint collusion attack, and fingerprint overwriting.


\subsection{Watermark trained model's output}

The category includes methods that focus on the output of a trained model instead of modifying training data or neural network parameters. Sebastian Szyller et al. \cite{szyller2019dawn} stated that existing watermark techniques focus on marking the models, which make their method vulnerable to model extraction attacks. Their method concentrate on watermarking the output instead of inputs, which means their approach watermarks a subset of API responses instead of watermarking a model. They introduced DAWN (Dynamic Adversarial Watermarking of Neural Networks) approach to use watermarking to deter model extraction IP theft. DAWN dynamically changes the responses for a small subset of queries (e.g., $1-2^{-64}$), and incurs a negligible loss of prediction accuracy (for instance, $0.03-0.5\%$).

Venugopal et al. \cite{venugopal2011watermarking} proposed a method to watermark the outputs of machine learning models, especially machine translation, to be distinguished from the human-generated productions. For watermarking, they used a random hashing operation by producing a fixed-length bit sequence. To detect the watermark, they used hashing operation on the outputs and found the bit sequences. The authors claimed that their methods are robust, uniformly distributed, and can distinguish between two generated data. Their results show that watermarking outputs have high recall and minimal quality degradation, and their producer can be identified.

\subsection{Attacks on Watermarking approaches}

The primary goal of watermarking a DNN scheme is to protect pre-trained models against intellectual property thefts and copyright infringement. However, the current algorithms suffer from some potential threats. There are many attacks designed to reveals the weaknesses of watermarking algorithms. In this section, we review some of the notable works proposed in this area.

Tianhao Wang et al. \cite{wang2019attacks} proposed an attack on watermarked DNNs. Their attacks are specially applied to watermarking approaches that modify the weights of the watermarked model, such as the model proposed by Uchida et al.\cite{uchida2017embedding}. Their attack is based on the variance of the model parameter distribution. They stated that the standard deviation of weights increased significantly during the process of watermark embedding. They proposed two different approaches to remove watermark: 1) Overwriting, which uses a single watermark and matrix in several epochs of training, and 2) Multi-embedding, which use different matrices and watermarks in each step of learning. L2 regularization is utilized to deflate the variance of the weights, which is the same as a non-watermarked model.

Shafieinejad et al. \cite{shafieinejad2019robustness} proposed three different attacks for backdoor-based watermarking: black-box, white-box, and property inference attack. In a black-box attack, they queried the watermarked model and used queries' output as labeled data. Their proposed white-box attack is inspired by fine-pruning techniques and has two subsection regularizations and fine-tuning. In the property inference attack, authors tried to detect whether a trained model is watermarked with backdoor approaches by extracting a feature vector from models and some parts of the training data. The experimental result shows that these three attacks could entirely remove backdoor-based watermarking.

Ryota Namba et al. \cite{namba2019robust} presented a novel attack method against query modification watermarking by changing the training data and finding a trigger set for watermark validation. Their attack has two steps, 1) key sample (trigger set) detection by measuring the changes after and before applying Autoencoder and 2) query modification for invalidating watermark. The authors also proposed a new watermarking method called \textit{exponential weighting}, which is robust against their attack method. Their proposed method recognized the model parameters that significantly affected prediction and increased their weight value exponentially. Finally, they demonstrated that their defense algorithm withstands under several different attacks, especially query and model modification.

\section{The Proposed Method} \label{ProposedMethod}

Text processing is one of the most common tasks in many machine learning domain with many applications in language translation, sentiment analysis, and spam filtering. Before using DNNs in text processing, many text processing tasks get stuck in a local optimum. Nowadays, DNNs significantly improve the performance of all text processing tasks. For instance, DNNs-based text classification plays a very critical role in understanding and analyzing the information itself. Since protecting the trained model in text processing is essential, preserving the trained models became a vital task for different industries and researchers. Hence, this paper proposes a framework for securely watermarking a textual DNN model.

\begin{figure}[t]
    \centering
    \includegraphics[width=0.65\textwidth]{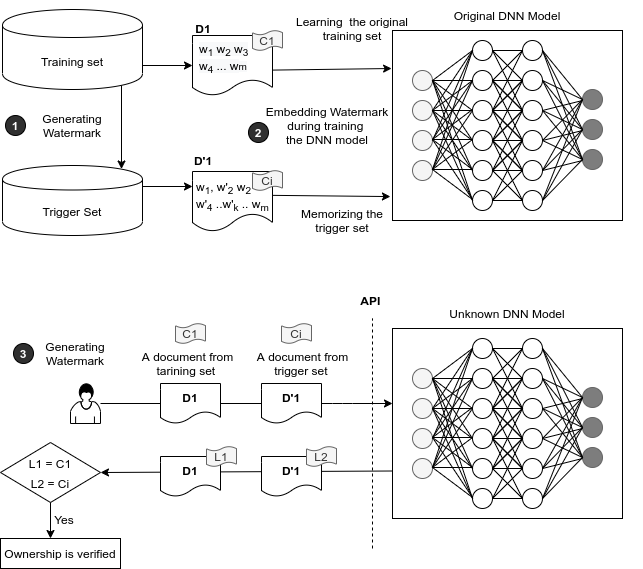}
    \caption{The workflow of the proposed DNN watermarking framework for textual data.}
    \label{fig:watermarkgen_EV}
\end{figure}

The three main components of the proposed method are watermark generation, watermark embedding, and watermark verification. In the watermark generation step, the content of some selected documents is changed and assigned with a new label. In the watermark embedding step, the watermarked documents are embedded in a trained model. These textual documents are called trigger sets, which are used in the watermark verification stage. In the watermark verification step, the model's ownership is examined using the trigger set generated in the first step. Figure \ref{fig:watermarkgen_EV} and the following step show the main workflow of the proposed DNN watermarking framework. The following sections describe each step in detail.

\begin{itemize}
\item Step 1: Randomly select $B$ samples from test data, and remove the stop words from them.
\item Step 2: Calculate the TF-IDF score for each word in all documents.
\item Step 3: For each selected document, randomly select one document from another class to exchange their words and producing a watermark record.
\item Step 4: Select K words of both documents with lowest TF-IDF score. 
\item Step 5: Exchange the selected words and swap the labels of two documents. 
\item Step 6: Insert the modified documents into the trigger set. 
\item Step 7: We repeat these steps (steps 3-6) until we meet all selected documents.
 \item Step 8: Combine the existing training set with the generated trigger set to form new training data.
\item Step 9: Train the DNN model with the new training data to achieve the proposed watermarked model.

\end{itemize}

 \begin{figure*}[h]
    \centering
    \includegraphics[width=0.99\textwidth]{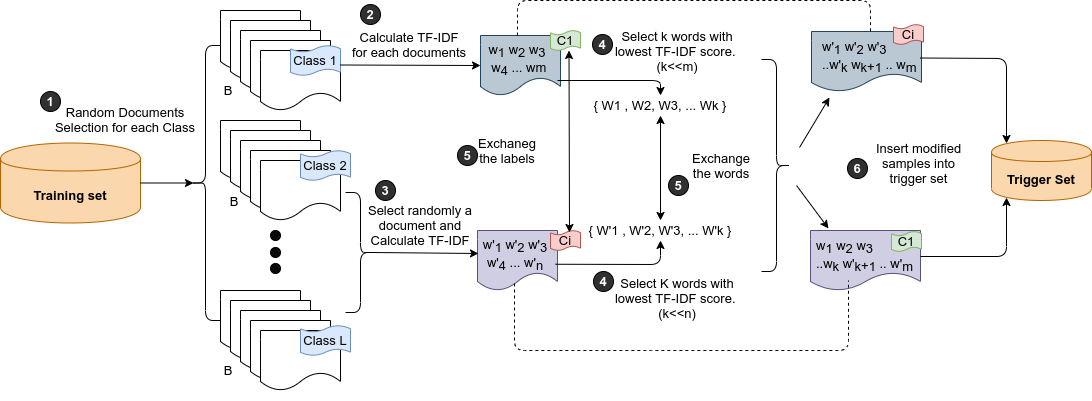}
    \caption{Steps of proposed watermark generation framework for textual data.}
    \label{fig:watermarkgen}
\end{figure*} 
 
\subsection{Watermark Generation}

The proposed algorithm in this paper generates a unique watermark to represent the owner's signature. The proposed algorithm securely watermarks the DNN model by utilizing an effective score called TF-IDF. This watermark is robust against several important attacks, such as reverse engineering methods. TF-IDF is a numerical statistic designed to rank essential words in a document based on their frequency. This score is a combination of Term Frequency (TF) \cite{luhn1957statistical} and Inverse Document Frequency (IDF) \cite{sparck1972statistical}. The TF-IDF score of a word $w$ of document $D$ can be calculated as below:

\begin{equation}
    S_{w,D} = tf_{w,D} \times \log \frac{N}{df_w},
\end{equation}

where $tf_{w_D}$ is the frequency of $w$ in $D$, $df_{w}$ is the number of documents contain $w$, and $N$ is the total number of documents. By using this score, we can rank all words in a document based on their importance. 

\begin{figure*}[h]
    \centering
    \begin{center}
    \begin{subfigure}[b]{0.45\textwidth}
    \includegraphics[width=0.99\textwidth]{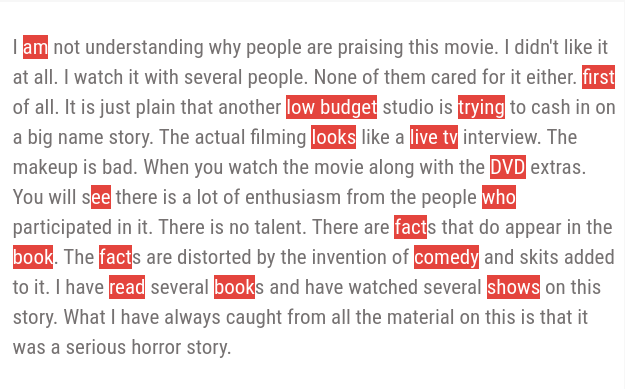}
    \caption{Original text}
    \end{subfigure}
    ~
    \begin{subfigure}[b]{0.45\textwidth}
    \includegraphics[width=0.99\textwidth]{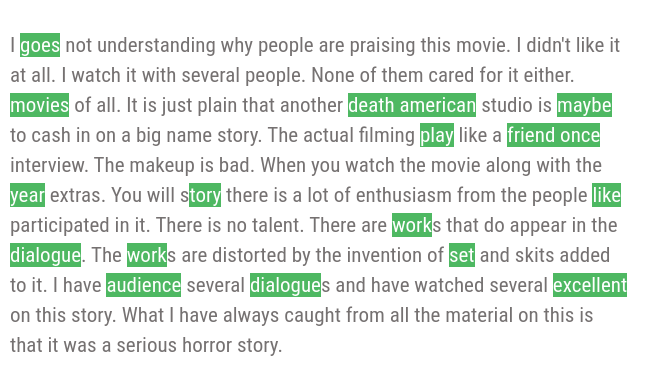}
    \caption{Modified text as a sample of trigger set. }
    \end{subfigure}
    \end{center}
    \caption{An example of watermark generation: 16 less important words are selected from original text(a), and randomly replaced with 16 less important words of another document to generate a sample of trigger set(b). }
    \label{fig:watermarkgen_example}
\end{figure*}

 Figure \ref{fig:watermarkgen} describes the proposed watermarking generation scheme. To generate a watermark, we randomly select $B$ documents for each class $C_i$ from the training set, $S_{C_i} = \{(D_j,Y_j) | Y_j=C_i \}_{j=1}^B \in {D}_{training}$. To create a fair and balanced trigger set, the number of samples selected from each class is equal. Then, we uniform the words by changing them to lowercase, and removing the punctuation and stop words. We calculate the TF-IDF score of each word, $w_m$, in those documents and sort all the words by their scores. 
Each document can be represented by its sorted words, $D_j = \{w_1, w_2, ..., w_m, w_{m+1},..., w_{n}\}$ where $n$ is the length of $D_j$ and TF-IDF$(w_m) \leq $ TF-IDF$(w_{m+1})$.  
For each document, $(D_j, Y_j) \in S_{C_i}$, we select $K$ words with the lowest TF-IDF values, $\{w_1, w_2, ..., w_K\} \in D_j$. 
We choose a sample from the other class randomly, $(D'_j,Y'_j) \in S_{C_{i'}}$ to exchange their lowest scored words, $D_j = \{w'_1, w'_2, .., w'_{K-1}, w'_K, w_{K+1}, ..., w_{n}\}$ and $D'_j = \{w_1, w_2, .., w_{K-1}, w_K,w'_{K+1}, .., w'_{n'}\}$. We exchange the labels of the two mentioned documents finally, $Y_j = C_{i'}$ and $Y'_j = C_i$. By doing these steps, we created watermarks that consist of a set of modified documents with the incorrect labels assigned to
them. This set is called a trigger set, $T = \{(D_j, Y'_j) , (D'_j, Y_j)\}_{j=1}^{B}$. Figure \ref{fig:watermarkgen_example} shows an instance of the original and watermarked document by the proposed method. The modified version can be used as the trigger set in the embedding stage. In this example, 16 words with the lowest score are selected from the original document and randomly replaced with the lowest score words extracted from a different document.


\subsection{Watermark Embedding}

Embedding a watermark into a DNN model can be done in one of the following three steps; 1) at the training time, 2) at the fine-tuning step, and 3) at the distillation step. We embed generated watermark data into the DNN model by inserting them in the proposed framework's the training step. We append the trigger set samples, list of modified documents, to the training data, and train the NN model to make it memorize the incorrectly assigned label to these samples. During the training step, our DNN tries to learn the labels of correct samples and memorize watermark examples. Therefore, the watermarks are embedded in the newly trained model. 
Algorithm \ref{WGE_algorithm} shows the pseudocode of the proposed DNN watermark generation and embedding approach.
After inserting watermark data into the DNN model, we must ensure that the model's performance did not decrease and that watermarked data are embedded in the model correctly. In Section \ref{ExperimentResults}, we define an experiment and evaluation metrics to prove that this expectation has been satisfied.

\begin{algorithm}[t]
\footnotesize
\SetAlgoLined
\DontPrintSemicolon
\SetKwInOut{Input}{input}
\SetKwInOut{Output}{output}
\Input{$D = \{D_j,Y_j\}^N_{j=1}$: Original Training Set with $N$ documents\;}
\Output{$M$: The trained DNN model \;
$\space$ $\space$ $\space$ $\space$ $\space$ $\space$ $\space$ $\space$ $\space$ $\space$ $T$: Trigger Set}
$T = \{\}$\;
 \For{each class, $C_i$}{
  $S_{C_i}=\{$select random $(D_j,Y_j)\in D\|Y_j=C_i\}^B_{j=1}$\;

  \For{each sample $(D_j,Y_j) \in S_{C_i} $}{
    Normalize $D_j$\;
    Calculate TF-IDF score for words in $D_j$\;
    Sort words of $D_j$ ascending, $\{w_m\}_{m=1}^n$ where $n$ is the length of $D_j$\;
    Select randomly $(D'_j,Y'_j) \in D$ where $Y'_j\ne Y_j$\;
    Normalize $D'_j$\;
    Calculate TF-IDF score for words in $D'_j$\;
    Sort words of $D'_j$ ascending, $\{w'_m\}_{m=1}^{n'}$ where $n'$ is the length of $D'_j$\;
    
    \For{$k = 1$ to $K$}{
        $w_k \Longleftrightarrow w'_k$
    }
    $D$ = $D - \{(D_j, Y_j), (D'_j, Y'_j)\}$\;
     $T$ = $T \cup \{(D_j, Y'_j), (D'_j, Y_j)\}$\;
  }
 }
 $M$ = Train ($D \cup T$)\;
 \Return{$M$ , $T$}
 \caption{Watermark Generation and Embedding}\label{WGE_algorithm}
\end{algorithm}

\subsection{Watermark Verification}

In watermark verification,  the ownership of a trained model should be verified. It means if an adversary creates a surrogate model without the owner's permission and provides an online API to service other users, we need a function to prove that his/her model is a surrogate. Therefore, we need to verify the ownership of this model. To verify the ownership, we send watermarked documents to the model. In case the predicted labels for the watermarked documents are the same as the expected labels (changed label), it can be verified that the trained model is derived from the correct model. It should be noted that the remote surrogate model may have modified to remove the watermark, so it is needed to send all documents in the trigger set to the remote API. Also, we must define a threshold, $\theta$, for ownership verification.

\section{Experiment Results}\label{ExperimentResults}

This section shows the results of the proposed watermarking algorithm for textual DNN models. The datasets used for analyses and the obtained result in watermark embedding and verification are described in the following.

\subsection{Dataset}

To evaluate the proposed watermarking approach, we use two well-known datasets in the text processing but for different problems; 

\begin{itemize}

  \item \textit{IMDB users' reviews}: IMDB users' reviews dataset, which is a well-known dataset in the Kaggle competition, is our first dataset. This dataset has two classes that show the polarity of each user's comment. It contains 25000 training samples and 25000 testing records. It is a standard benchmark for sentiment analysis problems. 

    \item \textit{HamSpam}: HamSpam as a well-known dataset in the spam detection area. This dataset is one of the most useful datasets in the spam detection task. It consists of 5728 email contents categorized into two main classes named \textit{Ham} and \textit{Spam}. It has 4360 samples for \textit{Ham} and 1368 samples for \textit{Spam}.
  
\end{itemize}

Both datasets are part of the Kaggle competition, and there are many methods developed based on these two corpora. Table \ref{db_tbl} summarizes the characteristics of selected real-world datasets. 

\begin{table}[h]
    \centering
    \small 
    \caption{Summary of data sets used in experiments}
    \label{db_tbl}
    \begin{tabular}{|c | c | c | c|}
    \cline{3-4}
         \multicolumn{2}{c|}{$ $} & \multirow{1}{*}{\textbf{IMDB}} & \multirow{1}{*}{\textbf{HamSpam}} \\ 
         \hline
         \multicolumn{2}{|c|}{Problem} & Sentiment Analysis & Spam Detection\\ \hline
        \multicolumn{2}{|c|}{Document length} & 231 & 19 \\\hline
        
        Number of  & train set & 24750 & 4317 \\\cline{2-4}
        positive  & test set & 250 & 43 \\ \cline{2-4} 
        samples & trigger set & 100 & 50 \\\hline
        
        Number of  & train set & 24750 & 1352 \\\cline{2-4}
        negative  & test set & 250 & 13 \\\cline{2-4} 
        samples & trigger set & 100 & 50 \\\hline
        
    \end{tabular}
\end{table}

\subsection{Evaluation Metrics} \label{Evaluation_Metrics}

There are a set of evaluation criteria that are widely used in the \textcolor{black}{literature} \cite{rouhani2018deepsigns,guo2018watermarking,quan2020watermarking} to show the effectiveness of a robust watermarking of DNNs:
\begin{enumerate}
    \item \textit{Fidelity:} The performance of the target model should not be noticeably decreased as a result of watermark embedding.
    \begin{equation}
        \mu(M(x; \theta^{*})) \approx \mu(M(x; \theta_0 )), s.t. \forall x \in X,   
    \end{equation}
    where $M(x; \theta_0)$ and $M(x; \theta^{*})$ denote the original DNN and watermarked model, respectively. $X$ is a set of documents used to train the original model, and $\mu(.)$ denotes any performance metrics such as accuracy, validation loss, or F1 score. 
    
    \item \textit{Integrity:} The ownership of the unmarked models must not be falsely claimed, i.e., the false alarm rate of watermark extraction should be minimum.

    \item \textit{Credibility:} The false negative rate of detecting embedded watermarks should be minimum. This is an essential requirement because the model owner needs to detect any misuse of her model with high probability. The watermarks can be effectively detected using the trigger set. 
    \begin{equation}
       \forall x_t \in T : A(x_t,\theta')=M(x_t,\theta^{*}) \iff A = M , \theta'= \theta^{*}
    \end{equation}
    where $A(.,\theta')$ is any useful DNN model for the same task and $M(.,\theta^{*})$ is the watermarked model. If the output of $A(.,\theta')$ is exactly the same as the output of $M(.,\theta^{*})$, then two models are the same and the claim of ownership is valid. Otherwise, fraudulent claims of ownership can be made. 
    \item \textit{Robustness:} Embedded watermark should be extracted after pruning, fine-tuning, and other model modifications. 
    \begin{equation}
        \mu(M(T; \theta^{*}+\varepsilon)) \approx \mu(M(T; \theta^{*} )),
    \end{equation}
    where $T$ denotes the trigger set and $\varepsilon$ is a small perturbation on the watermarked model's parameters ($\theta^*$).
    
    \item \textit{Efficiency:} The computational complexity of the embedding and extraction of the watermark should be insignificant.
    \item \textit{Security:} The robustness of the watermarking algorithm against attacks such as brute-force is essential. Leaving evidence in the targeted neural network can result in detecting or removing the watermark by a malicious actor.

\end{enumerate}

\subsection{Results}

Since the proposed method is based on word swapping, word selection strategy plays a significant role in the model's performance and robustness. We claim that the watermarked model is more robust and performs better when exchanging words with the lowest TF/IDF of two documents in the trigger set. To examine this hypothesis, we evaluate the proposed model by considering the following two strategies for exchanging the documents' words in the watermark generation stage.

\begin{table}[t]
\centering
\small
\caption{Fidelity score for watermarked and non-watermarked models.}
\label{table:fidelity}
\begin{tabular}{|c|lll|}\hline
       & Accuracy    & ASC  & DES  \\ \cline{2-4}
       \textbf{IMDB}  & Original Model   & 93.5\% & 93.5\% \\
       
        & Watermarked Model & 92.02\% & 91.8\% \\ \hline \hline
        & Accuracy    & ASC  & DES  \\\cline{2-4}
     \textbf{SpamHam}   & Original Model   & 98.3\% & 98.3\% \\
        & Watermarked Model& 97.5\% & 97.8\% \\\hline

\end{tabular}
\end{table}
\begin{itemize}
    \item Selecting the least important words (ASC): for this strategy, we sort the terms in ascending order according to their TF-IDF scores in both documents and then exchange the K=80 top words with each other.
    \item Picking the most important terms (DES): in this strategy, we replace the least significant word between two documents. So, we sort the phrases of each sample in descending order according to their TF-IDF values and replace the K=80 top words.
\end{itemize}

 We selected a different number of samples as a trigger set for HamSpam and IMDB datasets based on their sizes: 100 samples from HamSpam (B=50) and 200 samples from IMDB (B=100).

\begin{figure}[t]
    \begin{center}
    \begin{subfigure}[b]{0.48\textwidth}
    \includegraphics[width=0.99\textwidth]{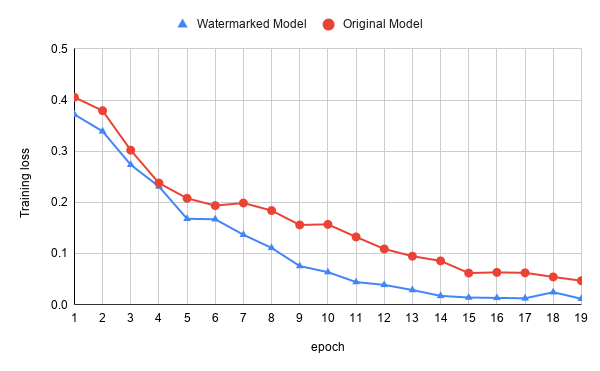}    \caption{Training loss}
    \end{subfigure}
    ~
    \begin{subfigure}[b]{0.48\textwidth}
    \includegraphics[width=0.99\textwidth]{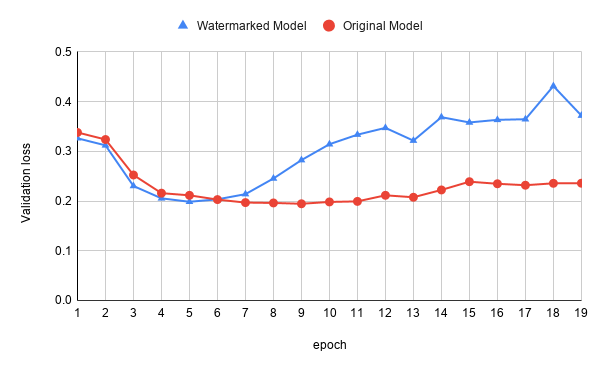}
    \caption{Validation Loss}
    \end{subfigure}
    \begin{subfigure}[b]{0.5\textwidth}
    \includegraphics[width=0.99\textwidth]{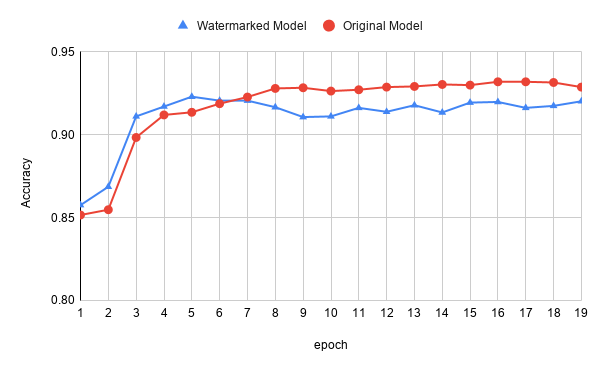}
    \caption{Accuracy}
    \end{subfigure}
    \end{center}
    
    \caption{Comparing Training loss, validation loss and accuracy of original and watermarked models considering IMDB dataset.}
    \label{Comparing_training_validation_accuracy_IMDB}
\end{figure}

\begin{table}[ht]
\centering
\footnotesize
\caption{The performance of original DNN model on \textit{IMDB} dataset in terms of training loss, validation loss, accuracy, precision, recall, and F1.}
\label{table:original_IMDB}
\begin{tabular}{|p{1cm}|p{1.8cm}p{2cm}p{1cm}p{1cm}p{1cm}p{1cm}|}
\hline
\multirow{1}{*}{epoch}  & Training loss & Validation loss& Accuracy & Precision & Recall & F1 \\
  \hline
1 & 0.406 & 0.338 & 0.852 & 0.82 & 0.892 & 0.855  \\ \rowcolor[HTML]{C0C0C0} \hline
2 & 0.379 & 0.324 & 0.855 & 0.817 & 0.906 & 0.859  \\
3 & 0.302 & 0.252 & 0.898 & 0.873 & 0.927 & 0.899  \\ \rowcolor[HTML]{C0C0C0}
4 & 0.238 & 0.216 & 0.912 & 0.891 & 0.935 & 0.912  \\
5 & 0.208 & 0.212 & 0.914 & 0.895 & 0.933 & 0.913  \\ \rowcolor[HTML]{C0C0C0}
6 & 0.194 & 0.203 & 0.919 & 0.907 & 0.929 & 0.918  \\
7 & 0.199 & 0.197 & 0.923 & 0.92 & 0.922 & 0.921  \\ \rowcolor[HTML]{C0C0C0}
8 & 0.184 & 0.196 & 0.928 & 0.917 & 0.937 & 0.927  \\
9 & 0.156 & 0.195 & 0.928 & 0.928 & 0.926 & 0.927  \\ \rowcolor[HTML]{C0C0C0}
10 & 0.157 & 0.198 & 0.926 & 0.917 & 0.935 & 0.925  \\
11 & 0.132 & 0.199 & 0.927 & 0.918 & 0.935 & 0.926  \\ \rowcolor[HTML]{C0C0C0}
12 & 0.109 & 0.212 & 0.929 & 0.917 & 0.939 & 0.928  \\
13 & 0.095 & 0.208 & 0.929 & 0.919 & 0.938 & 0.928  \\ \rowcolor[HTML]{C0C0C0}
14 & 0.085 & 0.222 & 0.93 & 0.926 & 0.932 & 0.929  \\
15 & 0.062 & 0.239 & 0.93 & 0.923 & 0.935 & 0.929  \\ \rowcolor[HTML]{C0C0C0}
16 & 0.063 & 0.235 & 0.932 & 0.919 & 0.944 & 0.931  \\
17 & 0.062 & 0.232 & 0.932 & 0.929 & 0.932 & 0.931  \\ \rowcolor[HTML]{C0C0C0}
18 & 0.054 & 0.236 & 0.932 & 0.93 & 0.93 & 0.93  \\
19 & 0.047 & 0.236 & 0.929 & 0.92 & 0.936 & 0.928  \\ \rowcolor[HTML]{C0C0C0}
\hline
\end{tabular}
\end{table}

\subsubsection{Fidelity} This metric shows the degree in which embedding a watermark affects the original model's performance. Table \ref{table:fidelity} shows the accuracy of watermarked and non-watermarked models for two datasets in two different strategies, selecting samples by ascending or descending TF-IDF scores. As the result shows, the accuracy of the watermarked model is very close to the non-watermarked model. So, we can claim that the proposed watermarking method in this research does not impair DNN performance. As Table \ref{table:fidelity} shows, both strategies embedded the watermark into the model without decreasing the performance.

\begin{figure}[t]
    \begin{center}
    \begin{subfigure}[b]{0.48\textwidth}
    \includegraphics[width=0.99\textwidth]{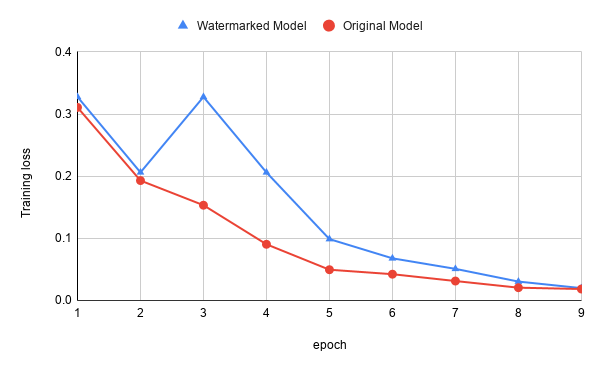}
    \caption{Training loss}
    \end{subfigure}
    ~
    \begin{subfigure}[b]{0.48\textwidth}
    \includegraphics[width=0.99\textwidth]{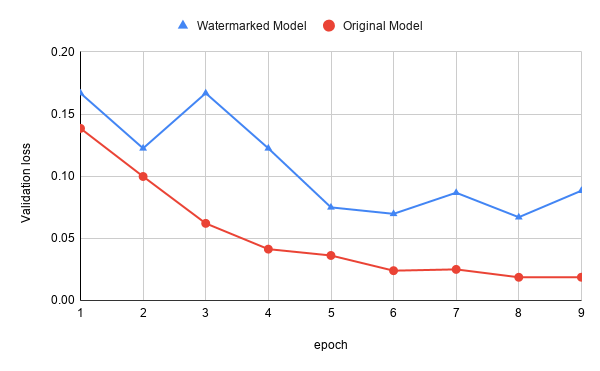}
    \caption{Validation Loss}
    \end{subfigure}
    \begin{subfigure}[b]{0.5\textwidth}
    \includegraphics[width=0.99\textwidth]{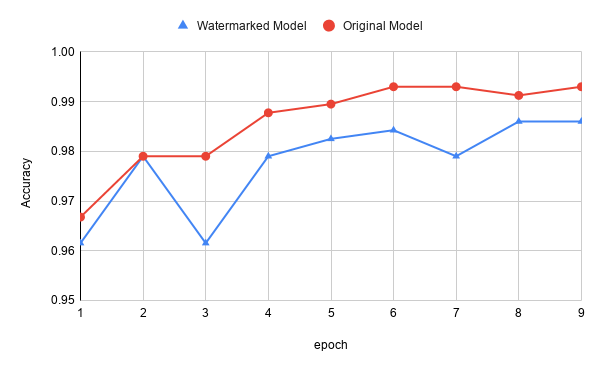}
    \caption{Accuracy}
    \end{subfigure}
    \end{center}
    
    \caption{Comparing training loss, validation loss and accuracy of original and watermarked models considering HamSpam dataset.}
    \label{Comparing_training_validation_accuracy_HamSpam}
\end{figure}

Figures \ref{Comparing_training_validation_accuracy_IMDB} and \ref{Comparing_training_validation_accuracy_HamSpam} illustrate the comparison between the original model and the ascending watermarking model in terms of training loss, validation loss and accuracy.
The proposed watermarking method does not impair the accuracy, along with not having any negative impact on precision, recall, and F1 score. Tables \ref{table:original_IMDB}, \ref{table:watermerked_IMDB}, \ref{table:Original_HamSpam} and \ref{table:Watermarked_HamSpam} show the performance of original DNN and watermarked DNN models on IMDB and HamSpam datasets, respectively. 

\begin{table}[t]
\centering
\footnotesize
\caption{The performance of watermarked DNN model on \textit{IMDB} dataset in terms of training loss, validation loss, accuracy, precision, recall, and F1.}
\label{table:watermerked_IMDB}
\begin{tabular}{|p{1cm}|p{1.8cm}p{2cm}p{1cm}p{1cm}p{1cm}p{1cm}|}
\hline
\multirow{1}{*}{epoch}  & Training loss & Validation loss& Accuracy & Precision & Recall & F1 \\
  \hline
1 & 0.372 & 0.326 & 0.858 & 0.859 & 0.858 & 0.858 \\  \rowcolor[HTML]{C0C0C0} \hline
2 & 0.339 & 0.312 & 0.869 & 0.892 & 0.841 & 0.866 \\
3 & 0.273 & 0.23 & 0.911 & 0.914 & 0.908 & 0.911 \\ \rowcolor[HTML]{C0C0C0}
4 & 0.231 & 0.206 & 0.917 & 0.906 & 0.932 & 0.919 \\
5 & 0.168 & 0.199 & 0.923 & 0.923 & 0.924 & 0.924 \\ \rowcolor[HTML]{C0C0C0}
6 & 0.167 & 0.204 & 0.921 & 0.913 & 0.931 & 0.922 \\
7 & 0.137 & 0.214 & 0.921 & 0.924 & 0.918 & 0.921 \\ \rowcolor[HTML]{C0C0C0}
8 & 0.111 & 0.246 & 0.917 & 0.903 & 0.935 & 0.919 \\
9 & 0.075 & 0.282 & 0.911 & 0.91 & 0.913 & 0.911 \\ \rowcolor[HTML]{C0C0C0}
10 & 0.063 & 0.314 & 0.911 & 0.932 & 0.888 & 0.909 \\
11 & 0.044 & 0.334 & 0.916 & 0.914 & 0.92 & 0.917 \\ \rowcolor[HTML]{C0C0C0}
12 & 0.038 & 0.347 & 0.914 & 0.921 & 0.907 & 0.914 \\
13 & 0.028 & 0.322 & 0.918 & 0.915 & 0.922 & 0.919 \\ \rowcolor[HTML]{C0C0C0}
14 & 0.017 & 0.369 & 0.913 & 0.906 & 0.924 & 0.915 \\
15 & 0.014 & 0.358 & 0.919 & 0.914 & 0.927 & 0.92 \\ \rowcolor[HTML]{C0C0C0}
16 & 0.013 & 0.363 & 0.92 & 0.92 & 0.921 & 0.92 \\
17 & 0.012 & 0.365 & 0.916 & 0.931 & 0.9 & 0.915 \\ \rowcolor[HTML]{C0C0C0}
18 & 0.024 & 0.431 & 0.917 & 0.887 & 0.957 & 0.921 \\
19 & 0.011 & 0.372 & 0.92 & 0.905 & 0.94 & 0.922 \\
\hline
\end{tabular}
\end{table}

\begin{table}[t]
\centering
\footnotesize
\caption{The performance of original DNN model on \textit{SpamHam} dataset in terms of training loss, validation loss, accuracy, precision, recall, and F1.}
\label{table:Original_HamSpam}
\begin{tabular}{|p{1cm}|p{1.8cm}p{2cm}p{1cm}p{1cm}p{1cm}p{1cm}|}
\hline
\multirow{1}{*}{epoch}  & Training loss & Validation loss& Accuracy & Precision & Recall & F1 \\
 \rowcolor[HTML]{C0C0C0} \hline
1 & 0.311 & 0.139 & 0.967 & 0.895 & 0.994 & 0.972 \\ \hline
2 & 0.193 & 0.1 & 0.979 & 0.938 & 0.987 & 0.977 \\ \rowcolor[HTML]{C0C0C0}
3 & 0.153 & 0.062 & 0.979 & 0.933 & 0.994 & 0.981 \\ 
4 & 0.09 & 0.041 & 0.988 & 0.962 & 0.994 & 0.987 \\ \rowcolor[HTML]{C0C0C0}
5 & 0.05 & 0.036 & 0.99 & 0.974 & 0.987 & 0.984 \\ 
6 & 0.042 & 0.024 & 0.993 & 0.987 & 0.987 & 0.987 \\ \rowcolor[HTML]{C0C0C0}
7 & 0.031 & 0.025 & 0.993 & 0.981 & 0.994 & 0.991 \\ 
8 & 0.021 & 0.019 & 0.991 & 0.981 & 0.987 & 0.986 \\ \rowcolor[HTML]{C0C0C0}
9 & 0.018 & 0.019 & 0.993 & 0.981 & 0.994 & 0.991 \\ \hline

\end{tabular}
\end{table}

\begin{table}[t]
\centering
\footnotesize
\caption{The performance of watermarked DNN model on \textit{SpamHam} dataset in terms of training loss, validation loss, accuracy, precision, recall, and F1.}
\label{table:Watermarked_HamSpam}
\begin{tabular}{|p{1cm}|p{1.8cm}p{2cm}p{1cm}p{1cm}p{1cm}p{1cm}|}
\hline
\multirow{1}{*}{epoch}  & Training loss & Validation loss& Accuracy & Precision & Recall & F1 \\
 \rowcolor[HTML]{C0C0C0} \hline
1 & 0.282 & 0.112 & 0.976 & 0.976 & 0.976 & 0.972 \\ \hline
2 & 0.181 & 0.096 & 0.977 & 0.977 & 0.977 & 0.977 \\ \rowcolor[HTML]{C0C0C0}
3 & 0.172 & 0.081 & 0.974 & 0.974 & 0.974 & 0.981 \\ 
4 & 0.097 & 0.072 & 0.972 & 0.972 & 0.972 & 0.987 \\ \rowcolor[HTML]{C0C0C0}
5 & 0.057 & 0.066 & 0.976 & 0.976 & 0.976 & 0.984 \\ 
6 & 0.042 & 0.086 & 0.972 & 0.972 & 0.972 & 0.987 \\ \rowcolor[HTML]{C0C0C0}
7 & 0.025 & 0.084 & 0.976 & 0.976 & 0.976 & 0.991 \\ 
8 & 0.012 & 0.085 & 0.976 & 0.976 & 0.976 & 0.986 \\ \rowcolor[HTML]{C0C0C0}
9 & 0.013 & 0.088 & 0.976 & 0.976 & 0.976 & 0.991 \\ \hline

\end{tabular}
\end{table}
\subsubsection{Credibility}

This metric illustrates how the trigger set can distinguish between the watermarked model and the original one. In other words, credibility measures how we can extract watermarked effectively. For calculating this measure, we query the model with documents in the trigger set. Table \ref{table:reliability} shows the accuracy of both watermarked and non-watermarked models to predict the class of trigger set items.

The watermark can not be obtained from the watermarked model when we exchange words with the high TF-IDF scores because the latent features extracted from DNN are correlated with these types of words. It shows that the words with high TF-IDF scores have a deniable role in extracting features. Therefore, we swap the words with the low TF-IDF value and Table \ref{table:reliability} presents the promising results.
\begin{table}[t]
\centering
\small
\caption{Credibility score for watermarked and non watermarked model.}
\label{table:reliability}
\begin{tabular}{|c|lll|}
\hline

        & Accuracy    & ASC  & DES  \\\cline{2-4}
      \textbf{IMDB}  & Original Model   & 10.5\% & 8.4\%  \\
       
        & Watermarked Model & 98.0\% & 54.2\% \\ \hline \hline
       & Accuracy    & ASC  & DES  \\\cline{2-4}
     \textbf{SpamHam}    & Original Model   & 12.8\% & 10.6\% \\
        & Watermarked Model& 88.3\% & 53.7\% \\\hline

        
\end{tabular}
\end{table}

As Table \ref{table:reliability} shows, we embed a watermark into the model and extract it accurately in the Ascending strategy. In the IMDB dataset, the accuracy of the original model on the trigger set is 10.5$\%$, while the accuracy of the watermarked model is 98.0$\%$. The accuracy of the original model and watermarked model on the trigger set for the HamSpam dataset are 12.8$\%$ and 88.3$\%$, respectively. These results indicate that the proposed approach is credible and reliable in watermarking a DNN model.

\subsubsection{Robustness}

\begin{figure*}[b]
    \begin{center}
    \begin{subfigure}[b]{0.47\textwidth}
    \includegraphics[width=0.99\textwidth]{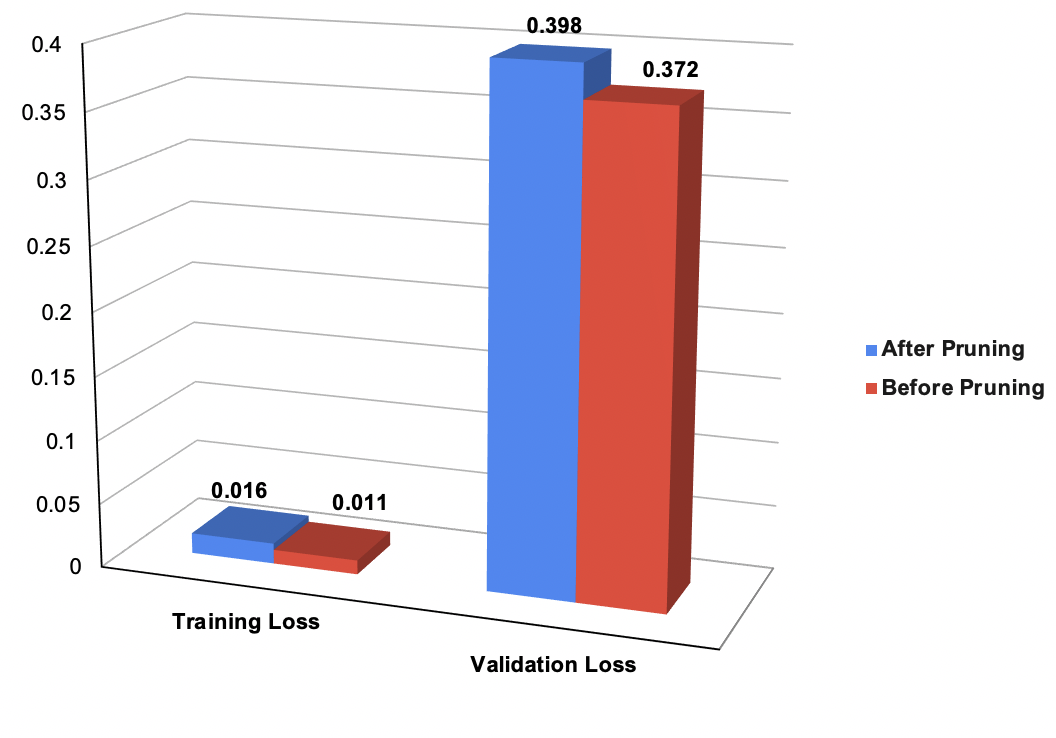}
    \caption{Training and Validation loss}
    \end{subfigure}
    \begin{subfigure}[b]{0.47\textwidth}
    \includegraphics[width=0.99\textwidth]{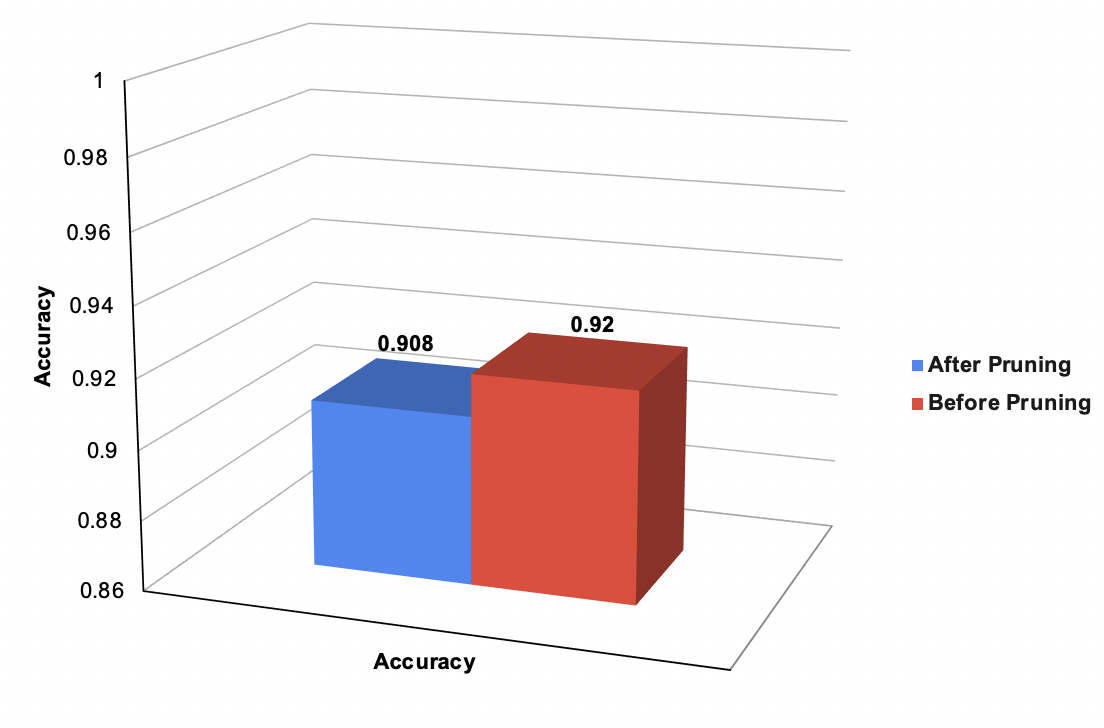}
    \caption{Accuracy}
    \end{subfigure}
    \end{center}
    
    \caption{Comparing training loss, validation loss and accuracy of watermarked model after applying parameter pruning. }
    \label{Robustness_parameter_pruning}
\end{figure*}

We use parameter pruning on the watermarked model for evaluating the Robustness of the watermarked model, which is trained on the IMDB dataset. In this stage,  the pruning approach is applied to sparsify the proposed watermarked model. A specific threshold is defined for a percentage of weights that are going to be replaced by zero. As Figure \ref{Robustness_parameter_pruning} shows, train loss, validation loss, and the accuracy of the watermarked model slightly decrease, but this amount is not noticeable. Thus, we can claim that our watermarked model is robust against parameter pruning, and parameter pruning did not impair our approach's performance.

\subsubsection{Efficiency}

The efficiency of the proposed method can be evaluated in each phase of our approach separately. Since watermark generation for a DNN is an offline process, so, it does not add any overhead to the prediction process of DNN. On the other hand, watermark extraction for the original and watermarked models are the same, and it takes time the same as simple querying to a DNN model. Therefore, evaluating the efficiency of the watermark embedding section is the most critical section. 

\begin{figure}[t]
    \begin{center}
    \begin{subfigure}[b]{0.47\textwidth}
    \includegraphics[width=0.99\textwidth]{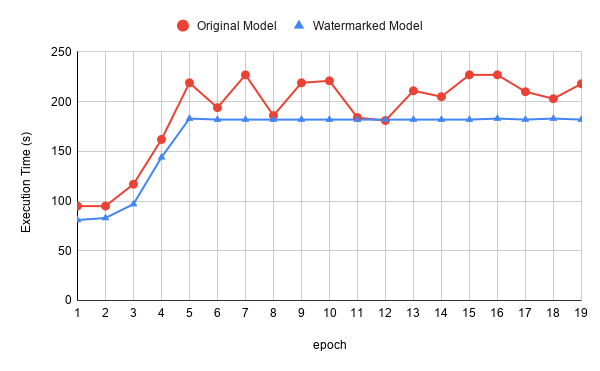}
    \caption{IMDB}
    \end{subfigure}
    \begin{subfigure}[b]{0.47\textwidth}
    \includegraphics[width=0.99\textwidth]{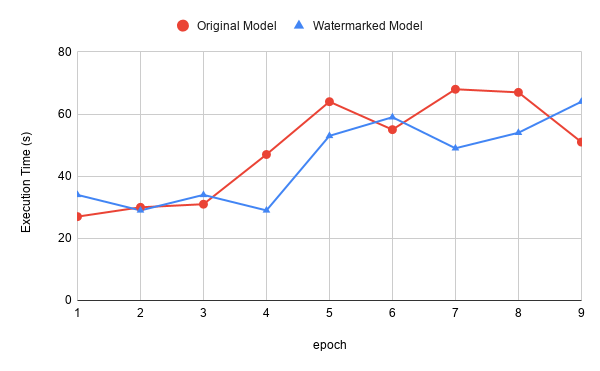}
    \caption{SpamHam}
    \end{subfigure}
    \end{center}
    \caption{Efficiency evaluation of watermark embedding in terms of execution time(s).}
    \label{fig:Execution_time}
\end{figure}

In this stage, the proposed watermark embedding schema's efficiency is evaluated by comparing the execution time of each epoch before and after the embedding procedure. The experiments in this stage are conducted using a machine with an Intel(R) Xeon(R) Silver 4114 CPU @ 2.20GHz CPU, 16 GB RAM and two Nvidia TITAN V 12 GB HBM2. Figure \ref{fig:Execution_time} illustrates the execution time comparison between the original model and the watermarked model. As Figure \ref{fig:Execution_time} shows embedding a watermark into a model slightly increases the execution time in contrast to the original model.

\subsubsection{Security}
As we explain in Section \ref{Evaluation_Metrics}, security measurement shows how a watermarked model is robust against the brute force attack. Since the primary focus of this research is textual data, the watermark input space is discrete and infinite. Therefore, embedded watermarks should be secure against brute-force attacks and also it is hard to guess or predict them.

\begin{figure}[t]
    \centering
    \includegraphics[width=0.6\textwidth]{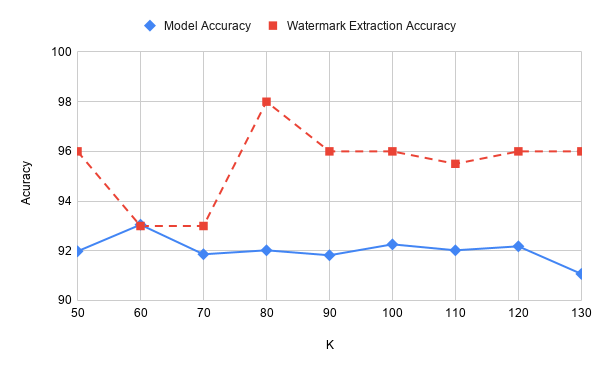}
    \caption{model accuracy and watermark extraction accuracy for different values for the $K$.}
    \label{fig:parametersetting}
\end{figure}

\begin{table}[t]
\centering
\footnotesize
\caption{The performance of watermarked DNN model on \textit{IMDB} dataset with different values of $k$ parameter in terms of training loss, validation loss, accuracy, precision, recall, and F1.}
\label{table:parameter_setting_IMDB}
\begin{tabular}{|p{1cm}|p{1.8cm}p{2cm}p{1cm}p{1cm}p{1cm}p{1cm}|}
\hline
\multirow{1}{*}{K}  & Training loss & Validation loss& Accuracy & Precision & Recall & F1 \\
 \rowcolor[HTML]{C0C0C0} \hline
50 & 0.014 & 0.396 & 0.92 & 0.938 & 0.893 & 0.915\\ \hline
60 & 0.01 & 0.311 & 0.931 & 0.93 & 0.932 & 0.931\\ \rowcolor[HTML]{C0C0C0}
70 & 0.016 & 0.345 & 0.919 & 0.917 & 0.917 & 0.917\\ 
80 & 0.011 & 0.372 & 0.92 & 0.905 & 0.94 & 0.922\\ \rowcolor[HTML]{C0C0C0}
90 & 0.011 & 0.417 & 0.918 & 0.905 & 0.931 & 0.926\\ 
100 & 0.009 & 0.345 & 0.923 & 0.929 & 0.918 & 0.923\\ \rowcolor[HTML]{C0C0C0}
110 & 0.009 & 0.389 & 0.92 & 0.931 & 0.905 & 0.918\\ 
120 & 0.013 & 0.344 & 0.922 & 0.926 & 0.907 & 0.916\\ \rowcolor[HTML]{C0C0C0}
130 & 0.011 & 0.445 & 0.911 & 0.935 & 0.882 & 0.907\\ 
\hline

\end{tabular}
\end{table}

In the watermark generation phase, $K$ words with the lowest TF-IDF score is selected to generate the trigger set. To experiment the effect of this parameter, we vary its value from 50 to 130. Table \ref{table:parameter_setting_IMDB} shows the performance of the watermarked DNN model on the IMDB dataset with different amounts of $K$ parameter in terms of training loss, validation loss, accuracy, precision, recall, and F1. The closest accuracy to the original model, which is $93.5\%$, is obtained when  $K=60$. However, watermark extraction accuracy is another important factor in selecting the best value for this parameter. Figure \ref{fig:parametersetting} illustrates both model accuracy and watermark extraction accuracy for the proposed approach by considering different quantities of $K$. As obtained results show, the watermarked model with $K=80$ achieves the best performance regarding both model accuracy and watermark extraction accuracy.

\section{Conclusion \& Future Works}\label{Conclusion}

Since collecting labeled data and providing powerful hardware for training a DNN model is costly, most scientists preferred to utilize a pre-trained model for different problems. Thus, securing a trained model becomes an essential task. In this paper, we applied the digital watermarking concept into the textual DNN models. We proposed an approach to protect a textual DNN model against copyright infringement and unauthorized redistribution. The proposed method did not decrease the original tasks' performance and is robust against different well-known attacks, such as parameter pruning. As experimental results demonstrated, the watermark can be extracted from the watermarked model accurately, which means the ownership of a trained model can be verified precisely.


The following potential hypothesis can be considered as the future work of this paper:
\begin{itemize}
    \item Analyzing the performance of the approach against all watermark attacks.   
    \item Generating a watermark with different methods
    \item Comparing the performance and robustness of the proposed algorithm with all existing frameworks
    \item Applying the proposed watermarking on various textual tasks such as phishing detection, machine translation and other textual deep neural network.
\end{itemize}

\bibliographystyle{ieeetr}
\bibliography{Watermarking}

\end{document}